\documentclass[aps,preprint,nofootinbib]{revtex4}

\usepackage{amsfonts}
\usepackage{amssymb}
\usepackage{amsmath}
\usepackage{latexsym}

\newcommand{\be}{\begin{equation}}
\newcommand{\ee}{\end{equation}}
\newcommand{\bea}{\begin{eqnarray}}
\newcommand{\eea}{\end{eqnarray}}
\newcommand{\benn}{\begin{eqnarray*}}
\newcommand{\eenn}{\end{eqnarray*}}

\def\bse{\begin{subequations}}%
\def\ese{\end{subequations}}%

\def\apjs{Astrophysical Journal Supplements}

\def\apss{Astrophysics and Space Science}

\def\prd{Physical Review D}

\def\aap{Astronomy and Astrophysics}

\def\plb{Physics Letters B}
\def\pla{Physics Letters A}

\begin{document}

\title{Generalized Slow-roll Inflation\\ in \\
Non-minimally Coupled Theories}
\author{A. Sava{\c s} Arapo{\u g}lu}
\email{arapoglu@itu.edu.tr}
\affiliation{Faculty of Science and Letters, Department of Physics, 34469 Maslak, Istanbul, Turkey}
\date{\today}

\begin{abstract}
The slow-roll field equations for the case of non-minimally coupled scalar fields are obtained in two ways: first using the direct generalization of slow-roll conditions in the minimal coupling case to non-minimal one; and, second, conformal transforming the slow-roll field equations in the Einstein frame to the Jordan frame and then applying the generalized slow-roll conditions.  We compare the difference of two methods in calculation of the spectral index, $n_s$, for a model example.  The second method seems to be more precise.  
\end{abstract}

\maketitle




\section{Introduction}

Inflation is the most plausible scenario providing not only the successful explanation of the horizon, flatness, and monopole problems of the standard big bang cosmology \citep{guth81, linde82, alb-stein82},  but also the primordial density fluctuations for the formation of the observed large-scale structure of the universe (references \citep{RMP97, PR99, RMP06, baumann09} for reviews). 

In inflationary universe models, it is supposed that the nearly exponential expansion of the universe is driven by a scalar field (called inflaton) which is assumed to be minimally coupled to the gravity and \textit{slowly} evolves in a nearly flat potential $V(\phi)$.  In the so-called ``slow-roll approximation'' \citep{sr94} the most slowly changing terms in the field equations are neglected which amounts to the approximation that the kinetic energy of the inflaton is considered to be much smaller than  the potential energy, that is, $\dot{\phi }^{2 }\ll V(\phi )$ and $\ddot{\phi }\ll H\dot{\phi }$.  The existence of inflationary attractors is necessary for slow-roll approximation to work.  The slow-roll single-inflaton field models predict almost scale-invariant density perturbations consistent with the observations of anisotropies in Cosmic Microwave Background (CMB). 

On the other hand, quantum field theory in curved spacetime necessitates a non-trivial coupling between the scalar field and the spacetime curvature even if they are absent in the classical theory.  Actually there are many other indications that the inflaton couples to the curvature of spacetime $R$ (summarized in a nice way in \citep{faraoni96}).  Therefore, it is reasonable to consider how the dynamics of the inflaton changes because of this non-minimal coupling.    In general, one expects that the coupling is of the form $\frac{1}{2}\xi \phi ^2 R$ with a constant $\xi $, but the quantum corrections may change this situation and the behaviour of renormalization group effective coupling $\xi$ becomes $\phi$ dependent also.  The various asymptotics of the coupling constant $\xi$ in quantum field theory in curved spacetime were studied in \citep{muta91}; recently in this direction the running of the non-minimal parameter $\xi$ is analyzed within the non-perturbative setting of the functional renormalization group (RG) \citep{shapiro15} and the inflationary parameters in the renormalization group improved $\phi^{4}$ theory at one-loop and two-loop levels are considered in \citep{odintsov14} and \citep{odintsov15}.  To cover all these effective models, then, one can consider a non-minimally coupled inflaton field with a general coupling function of the form $f(\phi)$.    

We are now currently in an era stated commonly as the `precision cosmology', implying that the observational data sharpens and this allows one to compare the models more precisely.  Inflationary models are examined and compared by the observations of Planck together with WMAP \citep{wmap13,planck14-22,planck14-16} and a joint analysis of BICEP2 \citep{bicep2planck15} via the inflationary parameters such as the spectral index $n_s$, the tensor-to-scalar ratio $r$, the running of the spectral index $\alpha=dn_s/d ln k$, and non-Gaussianity of the primordial perturbations.  Indeed, discriminating the various inflationary models through the calculation of these parameters in  both minimally and non-minimally coupled theories is an active research area.  Therefore, it is beneficial to consider and compare the calculation of these parameters in non-minimally coupled theories, and check whether there is any significant difference between minimal and non-minimal cases, considering the recent bunch of papers appearing in the literature about the subject.    

These parameters are obtained in the slow-roll approximation either considered directly in the \textit{Jordan Frame} (JF) through the ``generalized slow-roll'' approximation \citep{morris01, torres97}, or by performing a conformal transformation to the \textit{Einstein Frame} (EF) and using the usual definitions of slow-roll parameters \citep{sr94} in this frame; mostly the latter is preferred.  The existence of attractor behaviour in inflation with non-minimal coupling is also demonstrated in \citep{faraoni-gsr} which is necessary for this approximation to work.  The strong attractor inflationary behavior in multifield inflationary models is also shown in \citep{kaiser2014}.  Recently, Kallosh et al.\ \citep{kalloshlinde14} have shown that any inflationary model with a scalar-curvature non-minimal coupling asymptotes a universal attractor.

In this paper, we obtained the slow-roll field equations in JF in two ways: in the first method, we use directly the so-called `generalized' slow-roll conditions in JF, \citep{torres97, morris01}, and get the slow-roll field equations without any reference to EF; in the second method, we write the slow-roll field equations in EF, as they are originally suggested, and get the corresponding ones in the JF via conformal transformations and generalized slow-roll approximations together.  There is an interesting difference between the two methods:  although the slow-roll Friedmann equations coincide, the scalar field equations do not match exactly which leads to a difference in the calculation of, for example, the spectral index $n_s$.  It seems that the one derived from the EF slow-roll scalar field equation via conformal transformations provides a more precise value.  We exemplify this result with a simple model.

The plan of the paper is as follows:  In section II, we present the set-up and the background field equations, both in the JF and the EF.  In section III, we give two different ways of getting the slow-roll field equations in JF, and compare them in a simple model, showing explicitly their difference.  Section IV is devoted to concluding remarks.

\section{Set-up and Notation}
The action for the system of non-minimally coupled scalar field and gravity is
\be \label{action-JF}
S=\int d^{4}x\sqrt{-g}\left[ f(\phi )R-\frac{1}{2}(\nabla \phi )^{2}-V(\phi )\right] .
\ee
The field equations following from this action are
\bea 
f G_{\mu \nu }+(g_{\mu \nu }\square f-\nabla _{\mu }\nabla _{\nu }f)&=& \frac{1}{4} \nabla _{\mu }\phi \nabla _{\nu }\phi -\frac{1}{2}g_{\mu \nu }
\left[ \frac{1}{2}(\nabla \phi )^{2}+V(\phi )\right]  ,  \label{JFmetric-eom} \\
\square \phi &-& V^{\prime }+\frac{f^{\prime }}{f}\left[ 3\square f+\frac{1}{2}(\nabla \phi )^{2}+2V(\phi )\right] =0,\label{JFscalar-eom}
\eea
where a prime indicates $d/d\phi $.   
In flat Friedmann-Robertson-Walker spacetime, they become 
\bea 
H^2&=&\frac{1}{6f}\left ( \frac{1}{2}\dot{\phi }^{2}+V(\phi ) \right )-\frac{{f}'}{f}H\dot{\phi },\label{Jffriedmann}\\
\left ( 3\frac{{f}'^{2}}{f}+1 \right )(\ddot{\phi}+3H\dot{\phi })&+&\frac{{f}'}{f}({f}''+1)\dot{\phi }^{2}+f^{2}\frac{d}{d\phi }\left ( \frac{V}{f^{2}} \right )=0 \label{JFscalar-frw}.
\eea

A conformal transformation of the form
\bea 
\hat{g}_{\mu\nu}&=&\Omega^2(\phi)g_{\mu\nu},\label{ct}\\
\Omega^2(\phi)&\equiv &\frac{2}{M_{\text{Pl}}^{2}}  f(\phi) \label{cf},
\eea
brings the action into the Einstein-Hilbert form with a canonical scalar field
\be \label{action-EF}
S=\int d^{4}x\sqrt{-\hat{g}}\left[ \frac{M_{\text{Pl}}^{2}}{2}\hat{R}-\frac{1}{2}(\hat{\nabla} \hat{\phi} )^{2}-\hat{V}(\hat{\phi} )\right],
\ee
where the new (canonical) scalar field $\hat{\phi}$ is defined in terms of the JF scalar field $\phi$ through the relation
\be \label{scalar-reln}
\frac{d\hat{\phi}}{d\phi}\equiv M_{\text{Pl}}^{2}\sqrt{\frac{f+3f'^2}{2f^2}}.
\ee
The scalar field potential in EF is also defined as
\be \label{pot-reln}
\hat{V}(\hat{\phi})\equiv \frac{M_{\text{Pl}}^{4}V(\phi)}{4 f^2}=\frac{V(\phi)}{\Omega^4}.
\ee
Note that in Eq.(\ref{pot-reln}), the right-hand side is written in terms of the JF scalar field $\phi$.  To write $\hat{V}$ in terms of $\hat{\phi}$ one must use Eq.(\ref{scalar-reln}) to get $\hat{\phi}(\phi)$ and invert it to find $\phi(\hat{\phi})$ which is in principle possible but in general difficult.  The field equations following from Eq.(\ref{action-EF}) are
\bea
\hat{G}_{\mu \nu } &=&\frac{1}{M_{\text{Pl}}^{2}}  \left[ \hat{\nabla}_{\mu }\hat{\phi} \hat{\nabla}_{\nu }\hat{\phi} -\frac{1}{2}\hat{g}_{\mu \nu }(\hat{\nabla}\hat{\phi}
)^{2}-\hat{g}_{\mu \nu }\hat{V}(\hat{\phi} )\right] ,\label{EFmetric-eom} \\
\hat{\square}\hat{\phi} -\hat{V}^{\prime }(\hat{\phi} ) &=&0 \label{EFscalar-eom},
\eea
where the covariant derivatives are with respect to the metric $\hat{g}_{\mu\nu }$ and a prime indicates $d/d\hat{\phi} $.
In flat Friedmann-Robertson-Walker spacetime, they become
\bea
\hat{H}^{2}&=& \frac{1}{3 M_{\text{Pl}}^{2}}\left [ \frac{1}{2} \left ( \frac{d\hat{\phi }}{d\hat{t}} \right )^{2}+\hat{V}\right ], \label{EFfriedmann} \\
\frac{d^{2}\hat{\phi }}{d\hat{t}^{2}}&+&3\hat{H}\frac{d\hat{\phi }}{d\hat{t}}=- \frac{d}{d\hat{\phi }}\hat{V}.\label{EFscalar-frw}
\eea

\section{Slow-roll field equations in JF}
In non-minimal inflationary models the method followed mostly in the literature is to map the model in JF via conformal transformations to a model in EF, presumably due to the fact that the field equations  and the procedure to follow are simpler in EF.  The slow-roll parameters in EF are defined as usual,
\bea
\epsilon &\equiv & \frac{M_{\text{Pl}}^{2}}{2 }\left [ \frac{{\hat{V}}'(\hat{\phi })}{\hat{V}(\hat{\phi })} \right ]^{2},\label{srp1}\\
\eta &\equiv & M_{\text{Pl}}^{2}\left [ \frac{{\hat{V}}''(\hat{\phi })}{\hat{V}(\hat{\phi })} \right ],\label{srp2}\\
\zeta &\equiv & M_{\text{Pl}}^{2}\left [  \frac{{\hat{V}}'(\hat{\phi }){\hat{V}}'''(\hat{\phi })}{\hat{V}^{2}(\hat{\phi })}  \right ]^{1/2},\label{srp3}
\eea
where the prime denotes $d/d \hat{\phi}$.  To proceed in EF one has to write $\hat{V}$ in terms of the EF scalar field $\hat{\phi}$; but since in general it is difficult to find $\hat{\phi}$ in terms of $\phi$ in closed form, the generally preferred strategy is to express each quantity of interest in terms of the JF quantities.  The slow-roll parameters, for example, are to be evaluated at $\hat{\phi}_{\text{hc}}$ which is the value of $\hat{\phi}$ at which the scales of interest cross the horizon during the inflationary epoch.  Although the calculation of the value of the field at horizon-crossing is not an easy task in both frames, by assuming that the scales of interest cross the horizon after $N$ e-folding before the end of inflation, we can write
\be \label{scales-comp}
e^{N}\equiv \frac{\hat{a}(\hat{t}_\text{end})}{\hat{a}(\hat{t}_\text{hc})}=\frac{\Omega _\text{end}}{\Omega _\text{hc}}\frac{a(t_\text{end})}{a(t_\text{hc})},
\ee  
where $\phi_{\text{hc}}$ appearing in $\Omega$ is the value of the JF scalar corresponding to $\hat{\phi}_{\text{hc}}$.  This allows us to consider the slow-roll parameters, mapped back to the JF, at correct time \citep{kaiser95}.  Therefore we need slow-roll field equations and need to solve them to get $a(t)$ and $\phi(t)$ in JF.  We get the slow-roll field equations in two ways in this section: first we apply the generalized slow-roll approximations \citep{torres97,morris01,faraoni-gsr} to get the approximate field equations assuming the existence of inflationary attractors\citep{faraoni-gsr} in phase space.  Second we write the slow-roll field equations in EF, and express them in terms of JF variables by applying the conformal transformations Eq.(\ref{ct}) together with the generalized slow-roll conditions.  

\subsection{Slow-roll field equations in JF via generalized slow-roll conditions}
The dynamics of inflationary models with a single minimally coupled inflaton is considered in the ``slow-roll approximation'' \citep{sr94} which amounts to the assumptions that the inflaton evolves slowly in comparison to the Hubble rate, and that the kinetic energy of the inflaton is smaller than its potential energy.  These conditions are expressed in a compact way as $| \ddot{\phi } | \ll H | \dot{\phi } |\ll H^{2}| \phi | $ and $\dot{\phi }^{2}\ll |V(\phi )|$.  The generalization of these conditions to scalar-tensor theories with a coupling function $f(\phi)$ which has a sufficiently fast convergent Taylor expansion is 
\be \label{gsr}
| \ddot{f } | \ll H | \dot{f } |\ll H^{2}| f |,
\ee
first pointed out by Torres, \citep{torres97}.

Direct application of these conditions to the field equations in JF, Eqs.(\ref{Jffriedmann}) and (\ref{JFscalar-frw}), leads to the approximate field equations of the form
\bea
H^{2}&\simeq & \frac{1}{6f}V, \label{JF-appfriedmann1}\\
3H\dot{\phi }&\simeq& 2V\frac{{f}'}{f}- \phi {V}'. \label{JF-appscalar1}
\eea

\subsection{Slow-roll field equations in JF via those of EF } 
The slow-roll field equations in EF are obtained from Eqs.(\ref{EFfriedmann}) and (\ref{EFscalar-frw}) with the usual slow-roll approximations as
\bea
\hat{H}^{2}&\simeq & \frac{1}{3 M_{\text{Pl}}^{2}}\hat{V}, \label{EF-appfriedmann} \\
3\hat{H}\frac{d\hat{\phi }}{d\hat{t}}&\simeq &- \frac{d}{d\hat{\phi }}\hat{V}.\label{EF-appscalar}
\eea

Applying the conformal transformations Eq.(\ref{ct}) in connection with the generalized slow-roll conditions Eq.(\ref{gsr}) together, we get the slow-roll approximate field equations in JF.  The slow-roll approximated Friedmann equation Eq.(\ref{JF-appfriedmann1}) is exactly the same as the one obtained in the previous section but the slow-roll approximated scalar field equation becomes
\be \label{JF-appscalar2}
3H\dot{\phi }\left ( 1+3\frac{{f}'^{2}}{f} \right )\simeq 2V\frac{{f}'}{f}- \phi {V}'
\ee
which is different from Eq.(\ref{JF-appscalar1}) derived in the previous section.  

The difference between the scalar field equations in JF implies that the results of calculation of $\phi_{\text{hc}}$ are different, and thus $\hat{\phi}_{\text{hc}}$ and slow-roll parameters are different in turn. 

\subsection{Comparison through a simple example}
In this section we compare the Eqs.(\ref{JF-appscalar1}) and (\ref{JF-appscalar2}) through the calculation of the spectral index $n_s$ and check whether there is any meaningful difference between them in the light of the sharpening data of aforementioned observations.

To calculate the spectral index $n_s$, we must calculate slow-roll parameters in the EF and evaluate them at $\hat{\phi}_{\text{hc}}$.  But because of the possible technical difficulties in its calculation one can use the slow-roll scalar field equation to get the functional form of $\phi(t)$ and then Eq.(\ref{scales-comp}) to find the corresponding value in JF, for example, for $60$ e-foldings.  In this way we compare the possible difference between the slow-roll approximations we considered.

The spectral index to the second order in terms of these parameters is given by \citep{sr94,stewart93}
\be \label{spectralindex-2nd}
n_s=1-6\epsilon +2\eta +\frac{1}{3}(44-18c)\epsilon ^{2}+(4c-14)\epsilon \eta +\frac{2}{3}\eta ^{2}+\frac{1}{6}(13-3c)\zeta ^{2}
\ee
where $\gamma \simeq 0.577$ is the Euler's constant and $c\equiv 4(ln2+\gamma )\simeq 5.081$.
Therefore in obtaining $\phi_{\text{hc}}$ any difference in slow-roll field equations will make a difference in the spectral index obviously.  

We consider a specific model, the model of induced-gravity inflation\citep{kaiser94, kaiser94-b,spok84,turner85,lucchin86,fakir90,
dehnen95}, to compare the calculations, 
\be \label{action-example}
S=\int d^{4}x\sqrt{-g}\left[ \frac{1}{2}\xi \phi^2 R-\frac{1}{2}(\nabla \phi )^{2}-V(\phi )\right]
\ee 
together with the Ginzburg-Landau potential
\be
V(\phi)=\frac{\lambda}{4}(\phi^2-v^2)^2.
\ee
In this model, $\xi$ is the coupling strength of scalar field with curvature, and the non-minimal coupling makes the Planck mass a dynamical quantity through the relation $M_{\text{Pl}}=\sqrt{8\pi \xi}v$.  For this model, the new inflation initial conditions lead to the result for the spectral index to first order
\be
n_s \simeq 1-16\xi
\ee
if we use Eqs.(\ref{JF-appfriedmann1}) and (\ref{JF-appscalar1}), and give the result
\be
n_s \simeq 1-\frac{16\xi}{1+6\xi},
\ee
if we use Eqs. (\ref{JF-appfriedmann1}) and (\ref{JF-appscalar2}).  Although the difference between the two results differs by second order in the (small) parameter $\xi$, it is thought to be crucial because the inflationary theoretical predictions have now reached to level of second order in the slow-roll parameters.

The existence of a difference between the two approaches is interesting in that Morris \citep{morris14} shows that JF field equations, if expressed in terms of EF variables, agree with the EF field equations directly obtained from the EF action, provided that some consistency conditions are satisfied, and that these conditions are always met.  This result implies that the two frames are, at least, mathematically equivalent which in turn implies that one can work in one frame, if there is any advantage of simplicity over the other, and then can go to the other frame.  Further Kaiser showed that the spectral indices are the same in JF and EF \citep{kaiser-arxiv95}. The route that we follow here is in the reverse order: we obtained the approximate JF equations of motion from those of EF expressed in terms of JF variables and we compare it with the approximate equations of motion obtained directly in the JF.  The difference in the results does not seem to be because of the mathematical in-equivalence of the frames but stems from the fact that the slow-rolling approximation is a very critical issue and must be applied carefully for the non-minimal coupling case.  From our point of view the method that first writing the slow-roll field equations in EF and then expressing them in terms of JF variables together with the generalized slow-roll parameters seems to be safer and more precise.

The change in the scalar field equation can be expected on the ground that the conformal transformations themselves are dependent on the JF scalar field $\phi$ and that the `generalized' approximation directly in the JF cannot give exactly the same scalar equation obtained via conformal transformations from that of the EF.

\section{Conclusion}

In this work we have considered a general scalar-tensor theory of a single scalar field non-minimally coupled to the curvature with a general coupling function $f(\phi)$. The aim of considering such a system is application to single-field inflationary models in the slow-roll approximation.  The non-minimally coupled models in the inflationary context is an active research area and, with the proliferation of recent high-precision observations, is considered to be as important and viable as the standard minimally coupled models.  This is obvious in that there are many papers appearing frequently in the literature.  

Therefore we have thought that it is better once again (earlier in \citep{morris01}, \citep{torres97},  \citep{faraoni-gsr}) to consider the slow-roll field equations in such models by getting them in two different but related ways, comparing them and systematising the result.  A very careful analysis leads to a (presumably, though small) difference in the JF slow-roll scalar field equation of motion which may be meaningful in the light of current precise data.  This difference may cause important changes in the parameter space of non-minimal models studied currently.  The model example that we choose is in the form that is studied in the literature frequently.

The method that first getting the slow-roll field equations in the EF and then by applying conformal transformations in conjunction with the generalized slow-roll approximation seems to be safer and more precise than getting the JF approximate equations directly in the JF by applying the such generalized conditions.  One can expect that the conformal transformation of the slow-roll field equations in EF had to give the analogous equations in the JF; but having preformed the conformal transformations to JF, some additional terms are introduced which are able to be eliminated by the generalized conditions.  Thus we suggest to use Eqs.(\ref{JF-appfriedmann1}) and (\ref{JF-appscalar2}) as the more precise and correct set of slow-roll field equations in JF.  

Another issue that must be clarified is what could be the problem with the use of the generalized slow-roll approximation directly in the JF, i.e.\ Eqs.(\ref{JF-appfriedmann1}) and (\ref{JF-appscalar1}).  To answer this question one must remember that the slow-roll approximation is originally defined in the EF and it has a well-motivated physical content; direct generalization of this approach to the JF can possibly miss some underlying physical principles even if it seems mathematically proper. (An example of such a point in the literature is to disregard the necessity of attractor behaviour for such an approximation to work in the non-minimal context.)


\end{document}